# PHISHING DETECTION IN IMs USING DOMAIN ONTOLOGY AND CBA – AN INNOVATIVE RULE GENERATION APPROACH

Mohammad S. Qaseem[1] and A. Govardhan[2]

[1]Research Scholar, Dept. of CSE, ANU, Guntur, AP
[2]Director & Professor,, SIT, JNTUH, Hyderabad, Telangana

## ABSTRACT

*User ignorance towards the use of communication services like Instant Messengers, emails, websites, social networks etc. is becoming the biggest advantage for phishers. It is required to create technical awareness in users by educating them to create a phishing detection application which would generate phishing alerts for the user so that phishing messages are not ignored. The lack of basic security features to detect and prevent phishing has had a profound effect on the IM clients, as they lose their faith in e-banking and e-commerce transactions, which will have a disastrous impact on the corporate and banking sectors and businesses which rely heavily on the internet. Very little research contributions were available in for phishing detection in Instant messengers. A context based, dynamic and intelligent phishing detection methodology in IMs is proposed, to analyze and detect phishing in Instant Messages with relevance to domain ontology (OBIE) and utilizes the Classification based on Association (CBA) for generating phishing rules and alerting the victims. A PDS Monitoring system algorithm is used to identify the phishing activity during exchange of messages in IMs, with high ratio of precision and recall. The results have shown improvement by the increased percentage of precision and recall when compared to the existing methods.*

## KEYWORDS

*Instant Messengers, Phishing detection, Ontology based Information Extraction (OBIE), Classification based Association (CBA),triplet algorithm.*

## 1. INTRODUCTION

Instant messengers (IMs) have become an integral part of today's state of the art communication system with the latest gizmos, smartphones, tablets, laptops etc., becoming affordable and gaining popularity globally. Instant messaging initially started off as a chatting service between buddies, then gradually evolved into the most sought after and popular mode of communication. Ever since the Instant Messengers (IMs) were introduced viz., ICQ (I seek you), Internet Relay Chat (IRC), Microsoft's MSN Messenger, AOL Instant Messenger, Yahoo Messenger, Google Talk, etc. the user response they got was overwhelming [1]. Now IMs like Skype, WhatsApp and the likes of them have made their mark in the communication world with their popularity. With the enormous increase in scale of the IM's usage globally, it is highly desirous to make these services secure and reliable. It has been found that many security challenges do exist viz., most of the freeware IM programs do not have encryption capabilities, password management is not strong and have vulnerability to account spoofing and denial- of-service (DoS) attacks [1]. There was some reprieve when spam detection and filtering were introduced to the IMs [2]. But the most challenging issue being encountered is the detection and prevention of phishing in instant messengers as the existing fire walls available are not at all capable of detecting such attacks. With the large number of additional features available in the Instant Messengers, the potential areas for attack have also increased, as there are no concrete methodologies to counter Phishing attacks. This could lead to a profound effect on a phishing victim as he or she loses the trust in internet banking and e-commerce transactions by falling prey into disclosing confidential account details to the phisher's devious tricks. The consequences of such phishing attacks will have disastrous impact on the corporate and banking sectors and businesses which rely heavily on the internet, as the e-clients would lose their trust in the services owing to their vulnerabilities.

The term phishing is derived from fishing wherein fishers use a bait to do fishing. Here the attacker i.e. the phisher uses some socially engineered messages to perform phishing i.e. elicits personal information deceitfully from the unsuspecting victims through emails, instant messengers, websites, social networking medium etc. [3]. There were about 1,23,486 phishing attacks reported in the second half of 2012 and around 72,758 attacks reported worldwide in the first half of 2013 as analyzed by the Global Phishing survey 1H2013 [53] of the Anti-Phishing Work Group (APWG) [4]. Consider for instance a phisher tries to trick the unsuspecting victims to login to the fake website page and elicits confidential information, usually in online banking and e-commerce sectors though emails. In instant messengers, the phisher pretends to be a trust worthy chat mate and asks personal questions in order to figure out security details of bank accounts like passwords, codes, etc. from the unsuspecting victim. Even though active research has been done on phishing detection in websites, emails and social networking sites, research contributions found in Instant messengers are limited. There are no robust techniques developed to tackle the problem of phishing dynamically, although there are some techniques, where phishing detection is static in nature.

Data mining approaches were used to find the frequent phishing patterns and extract phishing rules on a larger scale in IMs but ultimately, the detected phishing words had larger percentage of false positives and false negatives. This was obvious as the entire detection process was content based and it failed to detect the threat activity pertaining to the phishing domain.

Even though in some IM systems, Association rule mining was used along with domain ontologies, which were applied right from information extraction and mapping them with pre-defined phishing rules to identify the phishing domains, still it could not detect the phishing words intelligently, as the main intention or *context* behind its usage could not be identified. But its performance was slightly better than those not using domain ontologies.

Avoiding ambiguity over the relevance of the identified phishing words is the main motive behind this work in order to enhance the instant messaging system performance. This could be done by identifying the *context* behind chatting messages in order to reduce the percentage of both false positives and false negatives.

The extracted domain and its context if mined together can generate more interesting phishing rules by using Classification Based on Association (CBA) rules, in order to generate instant phishing alerts dynamically for the victim, with the best possible performance i.e. maximum true positives percentage.

The paper is organized into 6 sections. Section 2 discusses the related work of prevalent phishing problems in Instant Messengers, the methodologies and approaches used to address the problem statement. Section 3 elaborates the architecture and design of the PDS methodology and discusses the architecture of OBIE for domain and context extraction using the Triplet Algorithms. Section 4 characterizes the phishing features so as to formulate its terminology apart from framing the pre-defined phishing rules based on key attributes. The application of Classification Based Association rules on the training datasets and test dataset to generate phishing rules is also discussed. Section 5 outlines the experimental setup to be established in the form of chat session transactions between chatters of IMs and addresses the implementation and testing of the Phishing detection system in IMs and evaluates the performance in terms of precision and recall. Section 6 summarizes and highlights the contributions of this research work in precise. It also provides directions for future enhancements in this research area.

## 2. RELATED WORK

Phishing detection in the past has not been dynamic i.e. pre-defined blacklisted phishing words were used for matching and could not detect the zero-hour phishing attacks, which is critical from the detection accuracy point of view. Also, when there is increase in the false positives the system would be more harmful rather than being useful, as the users will start ignoring the system warnings due to wrongly reported phishing alerts [3].

All the previously detected phishing words, URLs and IP addresses are listed and updated from time to time and stored as *blacklists*. These blacklists cannot provide protection from the zero-hour phishing attacks i.e. the phishing words or URLs which are not available in the blacklists previously. It took about 12 hours to detect and blacklist the new phishing words or URLs, which is a considerable delay owing to the fact that about two-thirds of the phishing attacks end within two hours [21]. Thus protection from zero-hour phishing attacks was not available. Google Safe Browsing API, DNS-based Blacklist (DNSBL), PhishNet are the examples of services using the blacklists. To overcome this problem, phishing heuristics i.e. experience based methods of problem solving were used like SpoofGuard, Phishguard, Phishwish, and Cantina [3].

Data mining techniques like association rule mining (ARM), classification and clustering have also been used to detect phishing through algorithms like Apriori, C4.5, Support Vector Machines (SVM), k-Nearest Neighbour (k-NN) and Density-Based Spatial Clustering of Applications with Noise (DBSCAN). Apriori algorithm is used for finding phishing patterns dynamically in Instant messengers from not only text messages but also from voice chatting, by integrating the IM system with speech recognition system [28]. This is purely a content based phishing detection system with increased FPs. Phishing detection in emails using C4.5 decision tree induction was done in [30] based on the highest information entropy, but it could not determine the phishing threat type.

Advanced classification method like Fuzzy Set Approach has been used efficiently by Maher et al. [32] to detect phishing websites in the E-banking sector. The intelligent fuzzy-based classification system also uses associative classification algorithms to detect phishing websites in the e-banking, and extracts the phishing features and classifies them into phishing rules with a layered structure [32].

Associative Classification (AC) uses CBA, CMAR, and CPAR to generate classifier rules, but the according to [16], the Classification based on associations (CBA) was more accurate than the C4.5 on large number of datasets. The iterative approach of Apriori is used by the CBA to create a classifier [16]. This AC approach has been used in [32] along with fuzzy based scheme for detection of phishing websites but nowhere for phishing detection in instant messengers.

In the above discussions phishing detection has been surveyed to a large extent in emails, URLs and websites but it has been observed that very less efforts have been made in the phishing detection in instant messengers. The purpose of studying the types of phishing approaches in emails, URLs and websites is to understand the extent of the research progress made in the three communication methods and how much progress remains to be achieved in instant messengers. The main observation here is that even though the zero-hour phishing detection was dynamic, but still there was no considerable decline in the false positives and the false negatives, as all these phishing detection approaches including phishing heuristics and data mining algorithms were content based methodologies.

Natural Language Processing (NLP) [47] techniques play a crucial role in overcoming the shortcomings mentioned above, by giving the advantage of utilizing the semantics of the messages exchanged between chatters in instant messaging. The Semantic Web uses the ontology and could incorporate the NLP to extract triplets i.e. subject, predicate and the object of the message under process, as a part of Ontology Based Information Extraction (OBIE). During the research study it has been observed that NLP has not been utilized to its full potential and at times it has only been used as an alternative to information extraction.

Another frame work which detects suspicious messages in Instant messengers and social networking sites using Ontology and Association Rule Mining (ARM) is used to detect the domain of the message keywords [33]. Pre-defined phishing rules have been considered which are to be mapped with the ontology generator through NLP, to obtain the domain of the suspected word. The user is not alerted instead the e-crime department is notified if any suspicious activity is detected [33]. The overall detection is content based and thus in spite of using the NLP, the **context** part is not explored, the use of ontology is restricted to information extraction and domain identification only. The overall performance does not reduce the false positives and false negatives.

# 3. DESIGN METHODOLOGY

Phishing detection in the past has not been dynamic i.e. pre-defined blacklisted phishing words

## 3.1. System Architecture

The architecture of the PDS Monitoring system consists of the following sub-systems:

- Instant Messaging system with clients and web browser.
- PDS monitoring system applying OBIE and CBA.
- Database (Message DB, Filter word DB, Ontology DB, Phishing rules DB and Phish word DB)

The Figure 1 shows architecture of the PDS monitoring system in Instant Messaging System (IMS), with the interrelationship among various subsystems in order to detect the phishing words based on context of the instant message and reports it to the victim client.

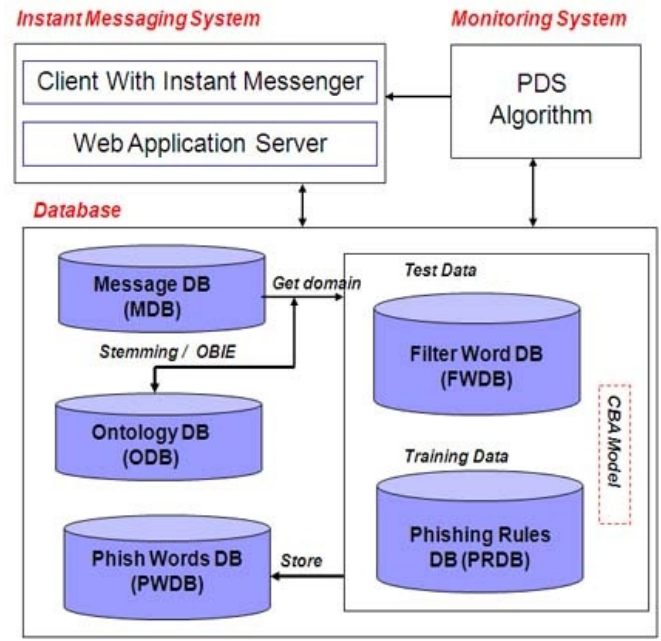

Figure 1.  Architecture of the PDS monitoring system in IMS.

The architectural sub-systems of the PDS Monitoring system as discussed in the following sub sections:

### 3.1.1 Instant Messaging System (IMS) with clients and web browser.

The IMS contains the following components:

- Various client machines with browsers.
- Web server containing the Instant Messaging Resources.
- IM Multiplexer for multiplexing the communications.
- Directory Server which helps in authenticating a client and contains the identities of the clients.
- Messaging server where the offline line messages are forwarded whenever the chatter starts the next session using SMTP.

### 3.1.2 PDS monitoring system applying OBIE and CBA.

The PDS monitoring system performs its task in two phases:

- Data pre-processing to extract domain and context using ontology (OBIE).
- Applying CBA and generate useful association rules

  In the first phase the PDS monitoring system captures and stores the instant messages in the Message DB (MDB) and then performs pre-processing on it through OBIE to identify the Domain and Context of the probable phishing word.

**Data pre-processing to extract domain and context using ontology (OBIE):**

The pre-processing includes removal of stop words and stemming.

***Stop Words Removal:*** From each instant message, stop words i.e. words that are not significant are removed, i.e. prepositions, conjunctions, articles, adjectives, adverbs, etc. [36]. Stop word examples are: from, into, in, for, while, a, an, the, that, these, those, under, over, about, although, how, what, when, who, whom, etc.  Indexing cannot be used on stop words and the removal of these words improves the efficiency of retrieval.

***Stemming*** is performed to reduce the derived or inflected words to their stem i.e. to its basic or root form. It is also referred to as conflation. Programs which perform stemming are called as stemmers or stemming algorithms [35]. Also stemming derives the stem morphologically from a completely suffixed word [34].   A stemmer identify the string "fishing", "fisher", "fishery", "fished" etc. based on the root "fish". The extracted keywords after pre-processing i.e. removing stop words and stemming are stored in the Filter Word DB (FWDB).

***Domain identification:*** After stemming is performed all the keywords obtained are in machine understandable form, which are to be mapped with the ontology to determine their respective domains through the OBIE. For the domain identification the first step is to use the Triplet Extraction Algorithm to extract the subject, predicate and object from the keywords. The NLP approach used here identifies a subject and maps it to a semantic class. It uses the predicate and object as name and value of the attribute respectively [12].

The next step is to identify the theme concept for which the concepts are to be extracted first. Concepts are the extracted tokens which are tagged as nouns and form the Concept set. The subjects identified above are populated into the subjectList. The concept which occurs the most number of times forms the MaxOccurConcepts. Identifying the Theme concept is done by performing the intersection of the three sets obtained [12].

Mathematically, Concept = [nouns]
subjectList = [subjects]
MaxOccurConcepts = [concepts]
ThemeConcept = concept $\cap$ subjectList $\cap$ MaxOccurConcepts [12]
Consider the following text passage:
"Hotel Taj Banjara at Banjara Hills offersv     excellent   facilities   and   accommodation. Comprising of 4 blocks and 68 deluxe rooms, Taj Banjara offers a pleasant stay."
Concepts = {Hotel, Taj Banjara, Banjara Hills, facilities, accommodation, blocks, rooms, stay}
   Mathematically, Concept = { nouns }
subjectList = {Taj Banjara}   subjectList = {subjects}
MaxOccurConcepts = {Taj Banjara (2)}
ThemeConcept = {Taj Banjara}
Themeconcept = Concept $\cap$ subjectList $\cap$ MaxOccurConcepts
Taj Banjara is identified as the theme concept

Now the Domain class can be identified by the mapping to the string Theme concept by using one of the two applicable rules: Explicit mention rule and the Implicit rule. Explicit mention rule suggests that the string which are the class names themselves. As far as the Implicit rule is concerned, it is used when there is no string match suggesting that the domain ontology lexicon. Domain ontologies are formed by experienced experts of the relevant domain. Semantic lexicons are created related to the respective domains, which is then used to identify the domain which the message content refers [12].

In the example the string Taj Banjara refers explicitly to **hotel**, and thus Taj Banjara belongs to hotel instance and the domain identified is **hotel**.

**Context identification:** This is an enhancement to the concepts discussed from [12] where in the task was limited to domain identification. The main contribution in this work is to identify the context or intention of the instant messages exchanged between chatters. The phishing detection may be misleading if phishing alerts are generated for harmless words, which are unintentional. This may leave a bad impression on the chatter to ignore such alerts again and again. Consider for example a message "Joe is so fond of chocolates. He would kill anybody for a bar of chocolate." Any phishing detection system would raise an alert for the word "kill", when actually its use is harmless. Extracting the true context is the main objective of this work apart from getting the least values for the false positives and false negatives.

In continuation to the previous discussion a new **predicateList** is introduced.

Concepts = {Joe, chocolate, bar} *Mathematically, Concept = {nouns}*
subjectList = {chocolate}     *subjectList = {subjects}*
predicateList = {fond, kill}
MaxOccurConcepts = {chocolate (2)}
ThemeConcept = {chocolate}
Themeconcept = Concept ∩ subjectList ∩ MaxOccurConcepts
"chocolate" is identified as the theme concept

It is found that domain for the theme concept "chocolate" does not exist in the domain ontology lexicon and comes under the Implicit rule. Thus a new Domain ontology for "eatables" is formed by experienced experts of the relevant domain using the classes, attributes and relation present in the ontology. New semantic lexicons (only a few a listed) have been created related to the "eatables" domain.

Eatables domain lexicon = {eat, dark chocolate, bar, nutty, lick, munch, kill, fond}

Note that, the words "kill" and "fond" are also included in the semantic lexicon of the domain "eatables", as they were a part of the predicateList. The new domain is mapped with the existing domains in the Pre-defined phishing rules DB (PRDB).

**If a match occurs Context = {harmful} or else Context = {harmless}.**

In the above example context = "harmless" in spite of the predicate or the verb being "kill".
The implementation of the context is possible with very simple sentences containing single values for the triplets.

The filtered keyword dataset in the FWDB, along with attribute values of Domain ontology and the Context forms the Test dataset for the generation of classification association rules (CAR) using CBA.

**Applying CBA and generate useful association rules:**

The Classifier Based Association (CBA) method explores all the associations between attribute values and their classes in the training dataset in order to build large classifiers. The training data set used here is the pre-defined phishing rules table shown in Table 1.

Table 1. Training data for the classifier based association rules *(snapshot)*

| S.no. | Keyword | Domain | Ontology context | Threshold value | Phishing word |
|-------|---------|--------|------------------|-----------------|---------------|
| 1 | Account no | Financial gain | Harmful | 3 | Yes |
| 2 | All caps | Account creation tips | Harmful | 5 | Yes |
| 3 | Pet name | Deceitful elicitation | Harmless | 5 | Yes |
| 4 | credit card detail | Financial gain | Harmful | 2 | Yes |
| 5 | Kids name | Identity access | Harmless | 1 | No |
| 6 | DOB | Financial gain | Harmless | 3 | Yes |
| 7 | Name | Not defined | Harmless | 5 | No |
| 8 | Whats App | Not defined | Harmless | 3 | No |
| 9 | Password | Fame and notoriety | Harmful | 1 | Yes |
| 10 | Hack | Not defined | Harmless | 1 | No |
| 11 | Supari | Not defined | Harmless | 1 | No |
| 12 | Spl char | Account creation tips | Harmful | 1 | Yes |
| 13 | *Image* | Not defined | Harmless | 5 | No |

The pre-defined phishing rules are framed by experts of the relevant domain having tremendous experience. The training dataset *(snapshot)* for the Classification Based Association rules is shown including the threshold values, Domain Ontology and its context. With the testing dataset rules are generated to predict the Phishing words and subsequently raising the phishing alert. If a new phishing word is identified based on threshold values based on frequently occurring words during chatting or a new phishing domain is discovered, these phishing words are appended to the Phishing Word DB (PWDB).

### 3.1.3  Databases

The backend used in the PDS monitoring system is Oracle 11g. There are total five database tables:

Message DB
Filter Word DB
Ontology DB
Phishing Rules DB
Phishing Words DB

## 3.2 System Workflow

The system workflow of the PDS monitoring algorithm initiates the steps to capture the phishing words from instant messages that are exchanged between the chatters through pre-defined phishing rules of Table 3.1 through Classification based on Association Rules (CBA) and Domain Ontology. The schematic illustration of the system workflow of the PDS monitoring algorithm is shown in Figure 2.

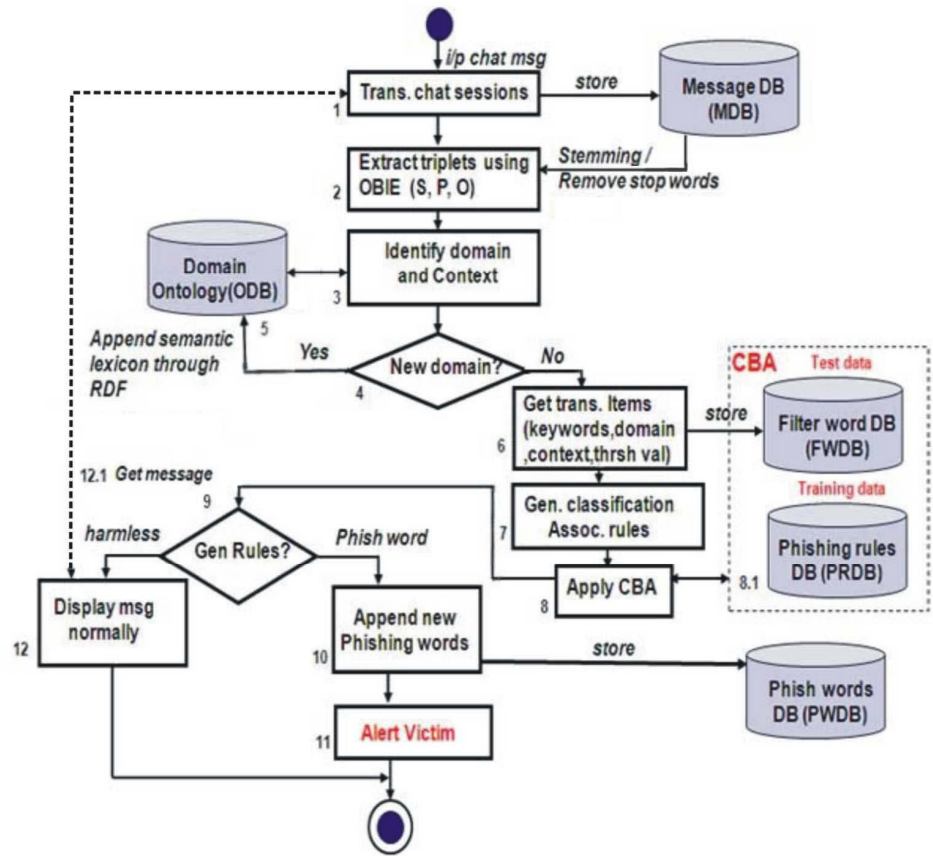

Figure 2 Activity diagram – System workflow of PDS Monitoring system

### 3.4 PDS Monitoring algorithm

***Input :*** Text messages of chatters
***Output :*** Phish word Alert to Victim
***Steps:***
 ***Start***

1. Store Messages in *MDB*
2. Apply OBIE           *//*Remove stop words, stemming, extract triplets
3. Get Domain ontology, Get  Context
4. If found newDomain then step 5 else step 6
5. Append semantic lexicon (ODB)    *//*through RDF
6. Store filtered words in *FWDB*   *//* These words along with domain and
                                        context attributes will be the Test data
                                        set. Update threshold value

7. Generate CAR rules        *// for the test data set*
8. Apply CBA        *//on the training data set pre-defined* (PRDB) and Test dataset to create classifier for generating phishing rules.
9. Generate Phishing rules
     if phish word = "YES" then step 10, step 11
     ELSE step 12
*10.* Append phish word to *(PWDB)*
11. Raise ALERT        *//to Victim Client*
*12.* Display message        *// original Message from (MDB).*
***Stop.***

The PDS monitoring algorithm elaborates the schematic work flow shown in the Figure 2 in a step by step manner.

This algorithm stores the instant messages in the MDB in step 1. Pre-processing is done on the messages through OBIE by removing stop words, performing stemming and extracting triplets (subject, predicate and object) in order to identify the Domain and Context in steps 2 and 3.

If a new domain is detected it is appended to the Ontology database (ODB) i.e. semantic lexicon, through RDF in step 5, else store the keywords in filter word DB (FWDB) in step 6.

In step 7, generation of Classification Association Rules (CAR) is done through test data formed from keywords in FWDB and the attributes (Domain, Context) obtained from step 3.

Apply Classification Based Association on the Training dataset i.e. the pre-defined phishing rule DB (PRDB) and Test dataset to create classifier for generating phishing rules in step 8.

In steps 9 to 12 the generation of Phishing rules is done and if phish word is found then the phish word is appended to the phishing word DB (PWDB) and an alert is raised on the victim's user interface, otherwise the message is displayed in its original form from the message database (MDB) on the Instant message interface.

## 3.5 The OBIE Architecture

Information extraction from the instant messages is required to be converted to machine understandable form in order to determine the domain and context of the instant message using Ontology Based Information Extraction (OBIE) component of the PDS monitoring system. This OBIE module initially reads the instant message keywords stored in the Filter Word DB (FWDB). The concept of Semantic Lexicon is used to identify the semantic domain and context for the keywords being processed through the domain inference module. The triplet algorithms are used here to assist in the identification of the Theme concept and the domain in the inference module [12].

The instance information is extracted by the instance extractor module. An RDF node is created and the ontology is updated using the Jena Apache APIs. Existing ontology editors like Protégé can be used for editing the ontology. The Lexicon extractor module contains rules in order to learn new lexicon symbols from the filtered keywords, and append them into the semantic lexicon using a set of heuristics to identify the relationship between lexical items and the existing semantic lexicon [12].The architecture for OBIE system is given below:

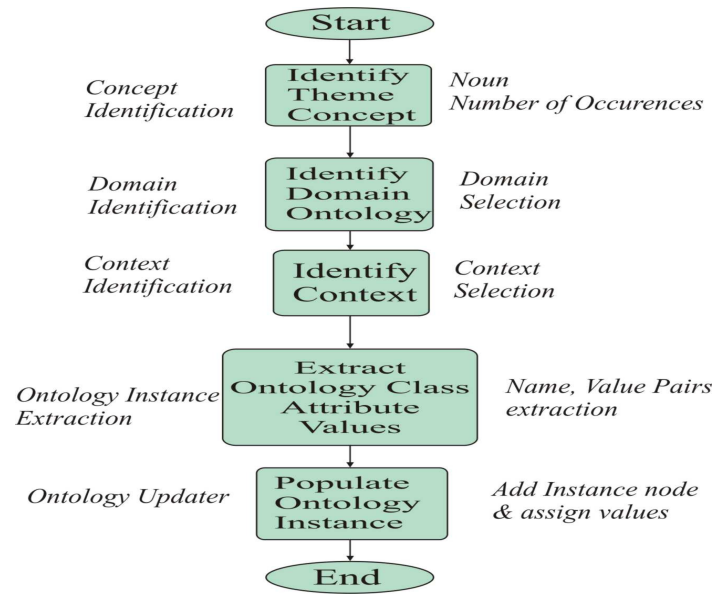

Figure 3 OBIE domain and concept extraction workflow

The figure 3.4 shows the workflow for the extraction of ontology instances.

## 3.6 Triplet algorithms

The Triplet Extraction Algorithms are used to extract the subject, predicate and object from the keywords for domain identification. The NLP approach used here identifies a subject and maps it to a semantic class. It uses the predicate and object as name and value of the attribute respectively [12].

The implementation of the triplet extraction algorithm mentioned in [37] has been done using the StanfordCoreNLP [38] Java library. The algorithms are briefly summarized for extraction of the subject, predicate and object.

*Algorithm* 1

*ExtractSubject (string)*
1. Perform a Breadth First Search (BFS) of the parse tree obtained by using StanfordCoreNLP library [38].
2. The NP subtree contains the subject, and it is the first Noun in the tree when traversed using BFS.

For complex noun compounds, the parse tree is used to extract all embedded Noun phrases (NP)[12] [40].

*Algorithm 2*
*ExtractPredicate (string)*

1. Perform a Depth First Search (DFS) of the VP subtree. The verb that is deepest in the tree is the predicate.

The parse tree is used to extract all embedded Verb phrases (VP)[12] [39][40].

*Algorithm 3*

***ExtractObject (string)***

1. Perform a search of the PP, ADJP subtree, and extract the first noun in the tree which is the object.
2.

The parse tree is used to extract all embedded preposition phrases (PP), and adjective phrases (ADJP)[12] [40]

For the previously used sentence "Novotel is located in Hyderabad", after applying the Triplet algorithm, the identified subject is Novotel, predicate is located, and object is Hyderabad. The theme concept identified here is Novotel and domain class extracted will be "hotel".

If the above sentence is expanded into a passage "Novotel, located at Banjara hills Road no. 1, offers memorable accommodation for the guests. It has 5 blocks and 67 luxury rooms. Novotel offers a decent ambience along with culinary specialties of the Deccan."

## 4. EXPERIMENTING THE PHISHING RULE GENERATION

Table 2 demonstrates the chatting between the two chatters.

| Chatter-1 | Chatter -2 |
|---|---|
| Hello do u hav any pets? | ya I hv 2 |
| Whats ur fav food | Its pizza |
| Who was ur fav teacher | Quite few |
| What is ur fav past time | Sudoku |
| What is ur lucky no | Guess …7 |

(a)    first session transaction

| Chatter -1 | Chatter -2 |
|---|---|
| Where d u live | At xyz |
| Which school d'ya study | The best in delhi, bbbbb |
| Whats ur age | 24, & urs |
| Elder to u by 2 yrs | Gud |
| Still, ur dob | 12-3-1990 |
| Ur dob same as my bro | I gotta go,. see ya |

(b)    second session transaction

| Chatter -1 | Chatter -2 |
|---|---|
| Was a tiring day at bank | Which bank |
| At xxxxx, and urs | Mine at yyyyyyyy |
| used internet banking | Ya |
| I haven't, any tips abt account creation | Try to use some caps |
| Thanks, bye | Bye |

(c)    third session transaction

| Chatter -1 | Chatter -2 |
|---|---|
| Hi, it says my password not strong | Try using ur kids name |
| It 'd we obvious | Use some spl chars |
| It lovely | Be careful to remember it |
| I'll get back to u | See ya |

(d)    fourth session transaction

| Chatter -1 | Chatter -2 |
|---|---|
| Hv u used debit card | Ya |
| Its asks for a 3 dgt code, where can I find it | Its on the back side of the card in Visa, mine is 654 |
| Thnks, yday my account created successfully | Any time pal |

(e) fifth session transaction

(f) sixth session transaction

......nth session transaction

Table 2 shows the chatting between the two chatters, where xyz, bbbb, xxxx & yyyyyy represents the place names.

## 4.1 Rule generation with CAR

Each chat session in Table 2 has a transaction id and the instant chatting messages are stored in the Message DB (MDB). Message pre-processing is performed by first removing the stop words and them stemming is performed in order to form the keywords to be stored in the Filter Word DB (FWDB).

Table 3 Filtered words stored in Filter word DB (FWDB)

| S.No. | Trans_id | Filtered keywords |
|---|---|---|
| 1 | T1 | Pet, favorite food, pizza, favorite teacher, favorite past time, Sudoku, lucky number |
| 2 | T2 | Day, bank, Citi bank, HSBC, internet bank, account create, cap |
| 3 | T3 | Live, Banjara hill, school, Delhi, DPS, favorite teacher, age, dob |
| 4 | T4 | Password, kid name, special character, careful, remember |
| 5 | T5 | Plan, kid, chocolate factory, kill, bar, Cadbury, time, favorite teacher , name |
| 6 | T6 | Debit card, digit, code, side, card, Visa, account, create, success, time, pal |

These filtered words are processed by the Ontology Based Information Extraction (OBIE) in order to obtain the keyword, Domain, Context, and threshold value for generating the test data as shown in Table 4.

The Association Classification rules for the three input components i.e. Ontology Domain, Context and threshold value are based on their values:

***Ontology Domain***: Financial gain, Account creation tips, Fame and notoriety, Deceitful elicitation, Identity access, Life threatening, not defined, and URL related.

***Context :*** Harmful and Harmless
***Threshold value :*** values 1to 5
The values of the ***Phishing word class*** are : YES, NO, and SPC i.e. is a phishing word, not a phishing word, and Suspicious word.
The Min_supp is 2% and min_conf is 60%

The CBA rules listed randomly as follows:
*Rule 1: If keyword = "account no" and domain = "financial gain" and ontology_context ="harmful" and threshold_value = "1" then Phishing_word = "YES"*
*Rule 4: If keyword = "credit card" and domain = "financial gain" and ontology_context ="harmful" and threshold_value = "1" then Phishing_word = "YES"*
*Rule 5: If keyword = "kid name" and domain = "identity_access" and ontology_context = "harmful" and threshold_value = "5" then Phishing_word = "SPC"*
*Rule 9: If keyword = "password" and domain = "fame_notoriety" and ontology_context = "harmful" and threshold_value = "1" then Phishing_word = "YES"*
*Rule 11: If keyword = "school" and domain = "not defined" and ontology_context = "harmless" and threshold_value = "1" then Phishing_word = "NO"*
*Rule 12: If keyword = "special character" and domain = "acc_creation_tips" and ontology_context = "harmful" and threshold_value = "1" then Phishing_word = "YES"*
*Rule 13: If keyword = "favorite food" and domain = "Ïdentity access" and ontology_context ="harmful" and threshold_value = "3" then Phishing_word = "YES"*
*Rule 15: If keyword = "debit card" and domain = "financial gain" and ontology_context ="harmful" and threshold_value = "1" then Phishing_word = "YES"*

The above phishing rules generated by the CBA will be used for training the test dataset.

A Phishing Word DB (PWDB) maintains the phishing words along with their respective domains, detected by applying the CBA through the PDS monitoring system. All new phishing words are appended to the PWDB table which comprises of 7 phishing domains. This table maps with the PRDB table for domain reference.

## 4.2 Applying CBA for Phishing Rule generation

The Test dataset along with the attributes and values i.e. Domain ontology, Context, and threshold values is shown in Table 4.

Table 4 Test data for the Classification Based on Association Rules (CBA)

| S.no. | Keyword | Domain | Context | Threshold value | Phishing word |
|-------|---------|--------|---------|-----------------|---------------|
| 1 | Favourite food | Identity access | Harmful | 3 | ? |
| 2 | Debit card | Financial gain | Harmful | 1 | ? |
| 3 | Lucky no | Acc_creation_tips | Harmful | 1 | ? |
| 4 | School | Identity access | Harmless | 1 | ? |
| 5 | Kid name | Identity access | Harmless | 1 | ? |
| 6 | Dob | Acc_creation_tips | Harmful | 1 | ? |
| 7 | Spl char | Acc_creation_tips | Harmful | 1 | ? |
| 8 | Password | Fame_noteriety | Harmful | 1 | ? |
| 9 | Kill | Not defined | Harmless | 1 | ? |
| 10 | Favourite teacher | Not defined | Harmless | 3 | ? |
| 11 | Code | Financial gain | Harmful | 1 | ? |
| 12 | Account | Financial gain | Harmful | 1 | ? |

In the above Table 4, the keywords along with their identified domains, contexts and threshold values are considered as the test data for Classification Association Rules (CAR). The classifier based on Classification based on Association (CBA) maps the test data with the training data i.e. the pre-defined phishing rules, to obtain the phishing rules to identify the phishing words.

The phishing rules generated using CBA on the above test data are shown as follows.

*If keyword = "favorite food" and domain = "Identity access" and ontology_context ="harmful" and threshold_value = "3" then Phishing_word = "YES"*

*If keyword = "debit card" and domain = "financial gain" and ontology_context ="harmful" and threshold_value = "1" then Phishing_word = "YES"*

*If keyword = "lucky no" and domain = "äcc_creation_tips" and ontology_context = "harmful" and threshold_value = "1" then Phishing_word = "YES"*

*If keyword = "school" and domain = "not defined" and ontology_context = "harmless" and threshold_value = "1" then Phishing_word = "NO"*

*If keyword = "kid name" and domain = "identity_access" and ontology_context = "harmful" and threshold_value = "1" then Phishing_word = "NO"*

*If keyword = "special character" and domain = "acc_creation_tips" and ontology_context = "harmful" and threshold_value = "1" then Phishing_word = "YES"*

*If keyword = "dob" and domain = "acc_creation_tips" and ontology_context = "harmful" and threshold_value = "1" then Phishing_word = "YES"*

*If keyword = "password" and domain = "fame_noteriety" and ontology_context = "harmful" and threshold_value = "1" then Phishing_word = "YES"*

*If keyword = "kill" and domain = "not defined" and ontology_context ="harmless" and threshold_value = "1" then Phishing_word = "NO"*

*If keyword = "favorite teacher" and domain = "identity_access" and ontology_context = "harmful" and threshold_value = "3" then Phishing_word = "SPC"*

*If keyword = "code" and domain = "financial gain" and ontology_context = "harmful" and threshold_value = "1" then Phishing_word = "YES"*

*If keyword = "account" and domain = "financial gain" and ontology_context = "harmful"*

*and threshold_value = "1" then Phishing_word = "YES"*

## 4.3 Raising Phishing Alerts

The following tabulations in Table 5 indicate the phishing words predicted by the CBA classifier through various colour indicators.

Phishing word: RED; Suspicious word: ORANGE; Normal word:BLACK

From the above mentioned colour indicators Suspicious words are predicted whenever the threshold value for a normal words >= 3.

Table 5 (a,b,c,d,e) Phishing words ALERT to the VICTIM chatter

| Chatter-1 | Chatter -2 |
|---|---|
| Hello do u hav any | ya I hv 2 |
| Whats ur fav food | Its pizza |
| Who was ur fav | Quite few |
| What is ur fav past | Sudoku |
| What is ur lucky no | Guess …7 |

(a) First session transaction

| Chatter -1 | Chatter -2 |
|---|---|
| Where d u live | At xyz |
| Which school d'va | The best in delhi. |
| Whats ur age | 24. & urs |
| Elder to u by 2 yrs | Gud |
| Still.. ur dob | 12-3-1990 |
| Ur dob same as my | I gotta go.. see ya |

(c) third session transaction

| Chatter -1 | Chatter -2 |
|---|---|
| Tiring day at bank | Which bank |
| At xxxxx, and urs | Mine at yyyyyyyy |
| used internet | Ya |
| I haven't, any tips abt account creation | Try to use some caps |
| Thanks, bye | Bye |

(b) second session transaction

| Chatter -1 | Chatter -2 |
|---|---|
| Hi, it says my password not | Try using ur kids name |
| It 'd we obvious | Use some spl chars |
| It lovely | Be careful to |
| I'll get back to u | See ya |

(d) fourth session transaction

| Chatter -1 | Chatter -2 |
|---|---|
| Hv u used debit card | Ya |
| Its asks for a 3 dgt code, where can I find it | Its on the back side of the card in Visa, mine is 654 |
| Thanks, yday my account created successfully | Any time pal |

(e) fifth session transaction

## 5. RESULTS AND OBSERVATIONS

**Sample Screenshot**

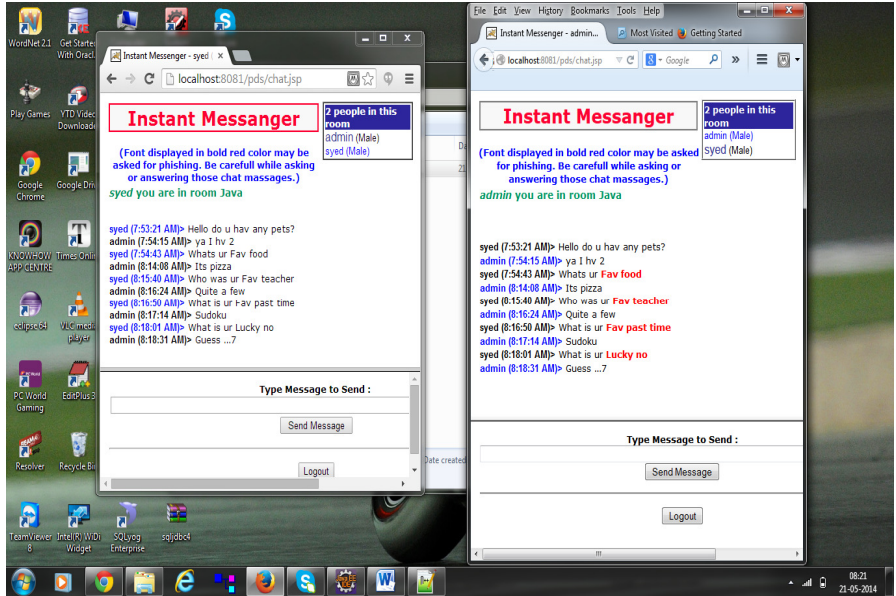

Figure 4  Screen shot Chat Session 1 with Phishing Alerts

**Performance**

The context oriented PDS Monitoring System has been tested on 110 transactions between the Chatters using domain ontology and generating improved rules through CBA to help detect Phishing words and generating ALERTS to the victim chatter.

Classifier performance is usually measured by accuracy, the percentage of correct predictions over the total number of predictions made. For this Precision and Recall are used:

**precision =     true positives / (true positives + false positives)**
i.e. the number of correctly predicted phish words.

**recall =        true positives / (true positives + false negatives)**

i.e. the total number of possible predicted phish words.

The total number of true positives (correctly identified phishing words) obtained out of 110 transactions were found to be 97.

The number of false positives (phishing words wrongly detected) out of 110 transactions were found to be 6.

Lastly the number of false negatives (phishing words which could not be detected) out of 110 transactions were 4.

There were 3 suspicious words detected which actually were phishing words.

**Precision = 97 / (97+6) = 94.17 %**
**Recall      = 97 / (97+4) = 96.03 %**

**Comparative study of the phishing detection systems in IMs**:
The comparative study shows that the CBA has a much better performance in terms of both precision and recall with respect to the existing systems mentioned above. Also the key feature being the context which provides the dynamism and intelligence element which provides an upper edge in the comparison.

Table 6. Comparative analysis

| Phishing detection Method | Domain Ontology | Based on | Precision | Recall |
|---|---|---|---|---|
| Deceptive Phishing Detection System in IM based on ARM | No | Content | 80.72% | 76.69% |
| Framework for surveillance of instant messages in IMs and social networking sites using data mining and ontology | Yes | Content | 85.67% | 84.36% |
| Context oriented PDS monitoring system based on Domain ontology and CBA in IMS | Yes | Context | 94.17% | 96.03% |

## 5. CONCLUSIONS

Instant Messaging Systems (IMS) generically cannot detect many deceptive phishing attacks; hence they are vulnerable for cyber frauds. To overcome this, a framework for detection of phishing attacks dynamically in IM from text messages using Classification Based Association rules and domain ontology is proposed. APDS monitors the user's psychology and predict the type of the detected phishing activity with an alert to the victim client.

## 6.1 Future Work

English is not the only medium of communication in a sub continent like India, it might be of multi lingual nature, the concept of translating and applying phishing detection is a future challenge. Words from other languages might be written in English which cannot be identified by any ontology based tool and may be ignored which in turn may turn out to be a phishing attempt. Images, sounds and videos are the formats which need a lot of work other than text formats where the vulnerability towards phishing may always be present.

The following issues and challenges are therefore identified where lot of research activities should be concentrated.

- Deceptive phishing messages are sent in any format other than textual (Images, Audio, Video), then they are not detected.

- Rules lack multilingual support for deceptive phishing detection.
- Issue with the interpretation of a message written in multiple languages.

**AUTHORS:**

Mohammad S. Qaseem , working as an Associate Professor, CSE Dept., Nizam Institute of Engineering and Technology, Hyderabad, India, is a research scholar pursuing part time Ph.D. in CSE from Acharya Nagarjuna University, Guntur, A.P., India, under the guidance of Dr. A. Govardhan, Professor and Director, SIT, JNTUH, Hyderabad.

Dr. A. Govardhan, is the Professor in CSE and Director, SIT, Executive Council Member, Jawaharlal Nehru Technological University Hyderabad (JNTUH), India. He served and held several Academic and Administrative positions including Director of Evaluation, Principal, Head of the Department, Chairman and Member of Boards of Studies and Students' Advisor. He has guided 28 Ph.D theses, 1 M.Phil, 125 M.Tech projects and he has published 305 research papers at International/National Journals/Conferences including *IEEE, ACM, Springer and Elsevier.*